\documentclass[superscriptaddress,preprintnumbers,floatfix,showpacs,showkeys,twocolumn]{revtex4}

\usepackage{epsfig}
\usepackage{amscd}
\usepackage{amsmath}
\usepackage{amssymb}
\usepackage{subfigure}

\begin{document}
\bibliographystyle{unsrt}


\title{Estimation of time-delayed mutual information and bias for
  irregularly and sparsely sampled time-series}

\author{D. J. Albers}  
\email{david.albers@dbmi.columbia.edu}
\affiliation{Department of  Biomedical Informatics, Columbia
  University, $622$ W $168^{th}$ St. VC-5, New York, NY 10032}

\author{George Hripcsak}  
\email{hripcsak@columbia.edu}
\affiliation{Department of  Biomedical Informatics, Columbia
  University, $622$ W $168^{th}$ St. VC-5, New York, NY 10032}

\date{\today}

\begin{abstract}
  A method to estimate the time-dependent correlation via an empirical
  bias estimate of the time-delayed mutual information for a
  time-series is proposed. In particular, the bias of the time-delayed
  mutual information is shown to often be equivalent to the mutual
  information between two distributions of points from the same system
  separated by infinite time.  Thus intuitively, estimation of the
  bias is reduced to estimation of the mutual information between
  distributions of data points separated by large time intervals. The
  proposed bias estimation techniques are shown to work for Lorenz
  equations data and glucose time series data of three patients from
  the Columbia University Medical Center database.
\end{abstract}



\preprint{arxiv.org/abs/nlin/0607XXX}

\maketitle


\section{Introduction}

In many experimental and computational contexts, experiments are
designed in such a way that the data that are collected are uniformly
and densely sampled in time, stationary (statistically), and of
sufficient length such that various time-series analysis can be
directly applied.  Nevertheless, there are an increasing number of
scientific situations where data collection is expensive, dangerous,
difficult and for which there is little that can be done to control
the sampling rates and lengths of the series.  Some examples could be
taken from astronomy \cite{non_uniform_astro}, atmospheric science
\cite{non_stat_aos1} \cite{non_stat_aos2}, geology
\cite{non_stat_geo1}, paleoclimatology
\cite{non_uniform_paleoclimatology}, seismology
\cite{non_uniform_seismology} geography \cite{non_stat_geography},
neuroscience \cite{non_stat_neuro1} \cite{non_stat_neuro2}
\cite{non_stat_neuro3}, epidemiology \cite{non_stat_epid}, genetics
\cite{non_uniform_genetics}, and medical data in general
\cite{non_stat_global_data} \cite{kantz_physiology}.

This paper primarily focuses on establishing a methodology for
confirming the existence of \emph{time-based} correlation for data
sets that may include highly non-uniform sampling rates and small
sample sets. To achieve this, we will use an information-theoretic
framework as an alternative to a more standard signal processing
framework because applying signal processing tools is difficult in the
context of time series with large gaps between measurements (e.g.,
spectral analysis for irregularly sampled points can be delicate to
implement \cite{non_uniform_ps_ieee_1988}
\cite{non_uniform_ps_review}).

More specifically, to circumvent the problem of erratic measurement
times, we use time-delayed mutual information \cite{kantz_book}
(TDMI), which does not rely on sampling rates, but rather on an
overall number of data points (we will explain this in detail in
section \ref{sec:coping_non_uniform_time}) used to estimate the TDMI
for a given time-separation. At a fundamental level, estimating the
TDMI requires estimating probability densities (PDF), and PDF
estimates fundamentally have two components, the PDF estimate and the
\emph{bias} (or noise floor) associated with the estimation of the
PDF. Therefore, relative to the information-theoretic context, the
problem of establishing time-dependent correlation can be reduced to
quantifying and estimating \emph{the bias of the statistical estimator
  used to estimate the probability density or mass functions}
(PDF/PMF) associated with a information-theoretic (mutual information)
calculation. (Note, for a more detailed review of nonlinear time
series analysis methods, see
Ref. \cite{santa_fe_time_series_prediction}; for more standard time
series analysis techniques see \cite{sprott_book} or
\cite{kantz_book}.)

The primary motivation for this work originates in work with
physiologic time series extracted from electronic health records
(EHR) \cite{ehr_use_1,ehr_use_2}. EHR data are never collected in a
controlled fashion, and thus are pathologically irregularly sampled in
time.  Nevertheless, EHR data will likely be the only medium- to
long-term source of population scale human health data that will ever
exist because human data can be too expensive, dangerous, and
intrusive to collect. Therefore, if we ever wish to understand
long-term physiologic trends in human beings, we must find a way to
use EHR data, and that means finding a way to cope with time series
whose values are mostly missing yet not missing at random. Here we
wish to establish a method to begin to quantify predictability of a
patient based on their physiologic time series. We hope to apply
predictability in at least three ways. First, in testing physiological
models on retrospective EHR, such as in glucose metabolism, we have
found that while it is sometimes impossible to distinguish two models
based comparing their outputs to raw data, sometimes models can be
distinguished by comparing their estimates of derived properties like
predictability \cite{pop_phys}. Second, in carrying out retrospective
research on EHR data, we often want to include only those patients
whose records are sufficiently complete, yet there is no simple
definition of completeness. We propose to use predictability as a
feature that correlates with completeness: a record that is
unpredictable across a wide range of variables may be so because it is
incomplete. And third, the predictability of a set of clinical or
physiological variables may be used as a feature to define a disease
or syndrome (or, for example, to subdivide the disease or
syndrome). Such a feature may add information beyond simple means,
variances, and simple characterizations of joint distributions.

\section{Nonlinear correlation in time: calculating the time-delayed
  mutual information}
\label{sec:tdmi}

\subsection{Non-uniformly sampled time-series: basic notation}
\label{sec:basic_notation}

Begin by defining a scalar time-series as a sequence of measurements
$x$ taken at time $t$ by $x_{t(i)}$ noting that $t$ denotes the
\emph{real time} that the measurement was taken, and $i$ denotes the
\emph{sequence time}, or sequential numeric index that denotes the
measurement's sequential location in the time series. Note that for a
discrete time series, $t$ is a natural number ($t \in N$), whereas for
a continuous time series, $t$ is a real number ($t \in R$). Denote
differences in real time by $\delta t = t(i) - t(j)$ where $i \geq
j$. Similarly, denote differences in sequence time by $\tau = i-j$
where $i \geq j$, noting that, regardless of the $\delta t$ between
sequential measurements, $i$ is always followed by $i+1$ (there are no
``missing'' measurements relative to sequence time).

\subsection{Time-delayed mutual information}
\label{sec:tdmi_defs}

Given two \emph{distributions} whose entropy \cite{cover_and_thomas} can be defined, the
amount that one distribution's information or differential entropy is
\emph{decreased} by knowing the other's distribution is known as the
\emph{mutual information} (MI) between the two distributions
\cite{shannon_math_theory_comm} \cite{cover_and_thomas}.  More
precisely, the MI between two distributions is defined by the
Kullback-Leibler divergence between the joint probabilities and the
product of the marginal probabilities. Here we denote the joint
probability distribution between a single variable measured at a time
$t$ and that same variable measured at a time $t - \delta t$ by
$p(x_t, x_{t-\delta t})$; similarly, the marginal distributions are
denoted $p(x_t)$ and $p(x_{t-\delta t})$. The \emph{time-delayed
  mutual information} (TDMI) between a variable measured at time $t$
and time $t-\delta t$ is then given by:
\begin{equation}
   I(X_t, X_{t-\delta t}) = \int p(x_t, x_{t-\delta t}) \log \frac{p(x_t,
    x_{t-\delta t})}{p(x_t) p(x_{t-\delta t})} dx_t dx_{t-\delta t}
\end{equation}
Thus, the TDMI measures the divergence between a product (orthogonal,
independent) distribution and a joint (correlated) distribution; note
that $I(X_t, X_{t-\delta}) = 0$ (up to the estimator bias) when $X_t$
and $X_{t-\delta}$ are independent. The maximum TDMI occurs for
$\delta t =0$ and is equal to the entropy of the time
series. Regardless of whether the TDMI is calculated for discrete
valued systems (using probability mass functions and the information
entropy), or continuously valued systems (using probability density
functions) the estimation of the TDMI depends entirely on the ability
to accurately estimate the PDF or PMF of the measured
distribution. Because establishing the existence of non-null MI
depends on the PDF/PMF estimation, establishing the existence of MI is
primarily a problem of establishing an rough upper bound on the bias
or noise floor (all MI above the bias is interpreted as real MI).

\subsection{Explicit estimation of probability mass and probability
  density functions}

There are roughly four different ways to estimate the PDF or PMF for
the mutual information calculation: (i) a standard histogram style
estimate \cite{pompe_MI}; (ii) an adaptive bin size histogram style
estimate \cite{fraser_swinney_MI}; (iii) a correlation function style
estimate \cite{grassberger_estimate_MI}; and (iv) a kernel density
style estimate (KDE) \cite{kde_MI_estimators}.  In this paper we
primarily use a KDE style estimate written for MATLAB
\cite{kde_matlab_I}. In the presence of a noisy signal, the authors
have observed no differences between the KDE and histogram style
estimators that lead to different interpretations of the
data. Nevertheless, the magnitude of the difference between the KDE
and histogram estimates does provide evidence for small sample size
effects. Thus, we will utilize a histogram estimator in some
situations to demonstrate the effects of the specific estimators. In
these situations, the histogram estimator is canonical, and has, as a
default, $256$ bins.


\subsection{Time-delayed mutual information for irregularly sampled or
  sparse data sets}
\label{sec:coping_non_uniform_time}

In the context of the TDMI, the data set used to generate the PDFs for
estimating the TDMI is generated by stepping though a time series and
collecting all pairs of points that are separated by some \emph{fixed}
time or time window. Thus, there is no real concept of missing points;
if a point does not have a corresponding point $\delta t$ in the past,
it is not included in the data set.  This means that, for a stationary
process, non-uniformity in the sampling rate effectively decreases the
sample size as it excludes points that do not have respective pairs
$\delta t$ in the past.  Said differently, given unlimited data string
lengths, uniform and highly non-uniformly sampling will yield
identical results. In a practical sense, a non-uniformly sampled time
series with $P$ pairs of points separated by $\delta t$ will render
the same as a shorter, uniformly sampled time series with $P$ pairs
of points separated by $\delta t$.  It is for this reason that TDMI is
a very natural measure of nonlinear correlation for systems that are
irregularly sampled in time. 

Because non-uniformity in sampling rates decreases the number of
available pairs of points to estimate the TDMI, nonuniform sampling is
really a data-sparsity problem, and relative to the statistical
estimators used to estimate the TDMI, a data-sparsity problem is
really a bias estimation problem.

\section{Bias estimation theory for information theoretic calculations}
\label{sec:bias_background}

Because the primary machines utilized for the TDMI calculation are PDF
(PMF) estimators, the primary sources of error or bias of the TDMI lie
with errors related to the PDF estimation.  In general, there are at
least three sources of bias related to PDF estimation techniques: (i)
bias due to the number of points present in the sample; (ii) bias due
to the resolution of the estimator such as the bandwidth or the number
of bins combined with the placement of the bin boundaries; and (iii)
bias due to the particulars of the data set, such as non-stationarity
or mixtures of statistically differing sources. Sources (i) and (ii)
can be at least accounted for; we briefly address in section
\ref{sec:MI_estimator_bias_explaination}.  Fundamentally, bias sources
related to (i) and (ii) are best represented through the classical
\emph{bias-variance tradeoff} present in histogram or KDE estimation;
we do not discuss that topic here. In contrast, bias related to
sources (iii), or bias due to the fundamental nature of complicated
data sources, is considerably more difficult to detect and quantify.
Nevertheless, at the end of this paper, we propose a method for
detecting the presence of non-estimator bias in some circumstances.

\subsection{Estimator bias calculations for entropy and mutual information}
\label{sec:MI_estimator_bias_explaination}

The bias associated with entropy and mutual information
calculation is always dependent on the particular PDF estimation
technique. Nevertheless, before we discuss estimator-specific TDMI
bias results, it is important to begin with the more
general framework within which analytic bias estimates are
calculated.

Begin by noting that the TDMI between two \emph{distributions} takes
the following parameters as arguments: the number of points, $N$; the
number of bins, $b$, (or bandwidth, $\frac{1}{b}$); and the real-time
separation, $\delta t$, yielding $I(N, b, \delta t)$. Assuming,
unrealistically, that for a given $\delta t$, $N \rightarrow \infty$
and $b \rightarrow \infty$ fast enough, then $I$ converges to some
value,
\begin{equation}
I(\infty, \infty, \delta t) = I_{\infty}(\delta t) 
\end{equation}
that has no bias and no variance. In lieu of having $N=\infty$ and
$\frac{1}{b}=\epsilon$ (where $\epsilon$ is an arbitrarily small
bandwidth), there will exist bias. In this situation, the result of
calculating $I$ will be:
\begin{equation}
I(N, b, \delta t) = I_{\infty} (\delta t) + B(N, b, \delta t)
\end{equation}
where $B(N, b, \delta t)$ is the bias of $I$ at $\delta t$ for a given
$N$ and $b$.  It is reasonably easy to calculate the biased mutual
information, $I(N, b, \delta t)$, so the key problem remaining is
the estimation of $B(N, b, \delta t)$.  

In the case where the statistical estimator used to estimate the PDFs
is the standard histogram, then the estimation of $B(N, b, \delta t)$
problem has been solved to second order in essence by Basharin
\cite{est_bias_histo_entropy}, who calculated the bias of the entropy
calculation, and in detail by Roulston (c.f. section $6$, on page $293$
of \cite{est_bias_histo_MI}).  (Higher order bias estimates of
entropies have been made for some algorithms in some circumstances,
c.f., \cite{grassberger_MI_error_estimates}.) The key qualitative
result of Ref. \cite{est_bias_histo_entropy} is, for a given $b$,
$\delta t$, and sufficiently large $N$, $B$ scales roughly as follows:
\begin{equation}
B(N, b, \delta t) \sim \frac{1}{N}
\end{equation}
Nevertheless, when the sample size is very small, the bias can be
influenced by the abundance of empty bins; or said differently, when
the bandwidth or bin resolution is too fine, it can increase the bias
substantially. Translating this to mathematics, the bias can be
expressed as:
\begin{equation}
B(N, b, \delta t) \sim \frac{A}{N}
\end{equation}
where $A$ is proportional to the amount of the support with no
measurements (e.g., the number of empty bins
\cite{est_bias_histo_MI}). Thus, when $N$ is small, for histogram
estimators, $A$ can be on the order of $N$ or larger; for KDE
estimators, because of their smoothing properties, the effects of the
empty support are translated into an over-weighting of empty
bins. Said simply, for small data sets, histogram estimators tend to
yield distributions with sharp peaks, while KDE estimators tend to
yield distributions that are closer to uniform distributions. Hence one of
the key issues addressed in this paper is the development of a
data-based estimation technique that can be applied to all PDF
estimation technique that is also fast and easy to use.


\section{Fixed-point method for estimating TDMI bias}
\label{sec:fp_estimation_method}

Before proposing a new method for estimating the bias of the TDMI
calculation, we briefly note the key features we hope to attain. We
want the bias estimator to be fast, easy to use, reliable and robust
to changing circumstances.  Additionally the bias estimation method
should apply to a wide variety of different estimators with different
bias properties; thus the method should be sensitive to, and thus
dependent on, the estimator being utilized. Finally, we want the bias
estimation technique to have the potential to depend on the data set
in the presence of non-estimator bias. In short, we want a bias
estimate for the TDMI based on the data set.  Because we are
calculating the univariate TDMI, we claim that very often the
infinite-time correlation will be the same as the bias of the
estimator, or:
\begin{equation}
\label{equation:infinite_bias}
      I(N, b, \infty) = B(N, b, \delta t)
\end{equation}
for all $\delta t$.  In essence, we are exploiting the fact that we
are interested in \emph{time-dependent} correlations, and thus we are
claiming that one can approximate $B(N, b, \delta t)$ with $I(N, b,
\infty)$ \emph{at least} for systems whose correlations decay to zero
in time. We further propose that $I(N, b, \infty)$ can be approximated
by removing or destroying the time-dependent information within the
time-series and then estimating $I(N, b, \delta t)$ repeatedly and
taking the average. Because the methods we propose rely, intuitively,
on approximating bias with the infinite-time correlation, we call this
bias method the fixed point bias, denoted as follows:
\begin{equation}
\label{equation:fp_bias_def}
B_{FP}(N, b, X) = \lim_{M \rightarrow \infty} \frac{1}{M} \sum_{i=1}^M
I(N, b, X, \delta t = i)
\end{equation}
where $X$ is the time-series (to be defined later) whose
time-dependent information has been removed. While it is possible to
construct examples where infinite-time probabilistic correlations
persist, in situations where the system is stationary, even when
nonlinear correlations in time persist, $B(N, b, \delta t)$ can still
be approximated when the time-dependent information has been
removed. In contrast, when the system is non-stationary, or when
multiple, possibly differing sources are aggregated, $B(N, b, \delta
t)$ will likely represent both estimator and non-estimator bias. 

In an effort to remove the time-dependent information from the time
series and thus approximate the estimator bias, we propose four
methods. Note that each method amounts to creating the two-dimensional
data set, $X$, referenced in Eq. \ref{equation:fp_bias_def}.  Further,
notice that the TDMI calculation acts as a filter selecting only the
\emph{pairs of points separated by a given $\delta t$}. For instance,
in the uniformly sampled discrete time case, a time series of length
$N$ will admit $N-\tau$ pairs of points for the PDF estimate. However,
in the situation where the time series is irregularly sampled, it is
possible that even for a very long time series there can be very few
points separated by a given $\tau$ or $\delta t$. For instance, in the
medical context, it is possible to have a patient whose time series
admits a large number of pairs of points separated by days and years
\emph{but very few points separated by a $\delta t$ on the order of
  months} (e.g., a patient with a chronic condition that acts up every
year or two and requires frequent measurements for a few days to
weeks.). Thus, for all the bias estimates below, it is important to
apply the method using only data collected on the $\delta t$ or $\tau$
time scale.

\textbf{Randomly permuted-data method:} Given the two-dimensional data
set used to estimate the TDMI for a given $\delta t$, randomly permute
or shuffle the data in one of the marginals \emph{without replacement}
(meaning, the permutation does not replace or change any of the data
points, it only changes the order of the data points).


To quantify an upper bound on the amount of time-based correlation
retained by randomly permuting the data, consider the \emph{mean
  distance} between the ordered and randomly permuted data sets. The
mean distance \emph{in time} between an ordered and randomly permuted
\emph{discrete-time} data set is given by:
\begin{equation}
\bar{\delta t} = \frac{2}{N^2} \sum_{i=1}^N \sum_{j=1}^N |i-j|
\end{equation}
After noticing a bit of symmetry, this equation can be reduced to:
\begin{align}
\bar{\delta t} &= \frac{2}{N^2} \sum_{i=1}^{N-1} \sum_{j}^{j<i} j \\
&=  \frac{1}{N^2} \sum_{i=1}^{N-1} i^2 + i \\
&= \frac{N}{3} - \frac{1}{3N} 
\end{align}
which, by the time $N=5$, can be approximated by $N/3$.  Note that
when time is continuous, and integrals replace sums, one arrives at
$N/3$ explicitly. Thus, the \emph{average} term by term time
correlation will be on the order of $N/3$.  Nevertheless, it is
important to keep in mind that while the \emph{average} time
correlation is $N/3$, the \emph{time-based pairings have been
  randomized}, and thus when integrating, considerably more time-based
correlation information has been lost --- hence the statement that
$N/3$ is a upper bound on the length of the $\delta t$ used to
estimate the correlation fixed point. It is worth noting that for in
many cases (e.g., in the context of many data points), randomly
permuting the data will not be a rigorous upper bound on the bias. For
instance, block-bootstrap methods, which retain autocorrelation
structure, will, depending on the circumstance, yield an even more
conservative estimate of the TDMI bias.


Recall again that we want the bias estimation method to be applicable
in circumstances where $N$ is small.  However, for small $N$, the
permuted data method can have two issues that require quantification:
(i) the effect of having a relatively short time-series which can lead
to the inability to approximate the $\delta t = \infty$ fixed point as
discussed in the previous paragraph (note that bootstrapping won't
help when all the points are too correlated); and (ii), the effect of
small sample size as related to the bandwidth/bin-width of the
estimator. Thus, the following bias estimation techniques are aimed at
addressing and quantifying both of these potential issues.

\textbf{Uniform-mixed method:} Given the two-dimensional data set used
to estimate the TDMI for a given $\delta t$, replace one of the
marginals with uniformly distributed random numbers.  Thus, with
respect to PDF approximation, this method preserves one marginal (the
raw data string) and replaces the other marginal with a uniform random
number.  If only the mean of the source-data is known, this method
will maximize the entropy of the sample by weighting all correlations
between subsets equally.

\textbf{Gaussian-mixed method:} Given the two-dimensional
data set used to estimate the TDMI for given $\delta t$, replace one of the
marginals with normally distributed random numbers with a given mean
and variance.  With respect to PDF approximation, this method again
preserves one marginal while replacing the other marginal with a
Gaussian random number.  If the mean and variance of the source-data
are known and taken into account, this method will maximize entropy.
In this work, we will always use mean zero and variance one.

\textbf{Accept-Reject generated random variable mixed method:} Given
the two-dimensional data set used to estimate the TDMI for a given
$\delta t$, replace one of the marginals with random numbers generated
by a distribution that approximates (using the accept-reject or like
method) the non-ordered \emph{distribution} of the original data-set.
This method is meant to fabricate the source-data most closely by
generating a set of random numbers with the same distribution as the
data.  This method will be the most computationally intensive of the
four methods, and for short or sparse data sets where it is meant to
excel, it will resemble the uniform-mixed method results.  In this
paper we will employ the standard accept-reject PDF fitting algorithm
with two different criteria, a strong fitting criterion (denoted
AR-mixed strong) and a weak fitting criterion (denoted AR-mixed weak).

\textbf{KDE-Histogram bin effect method:} Given the two-dimensional
data set used to estimate the TDMI for a given $\delta t$ and any of
the above methods for estimating bias, compare the TDMI fixed point
for a KDE estimator and a histogram estimator with roughly equivalent
bandwidth/bin-sizes. Because the smoothing property of the KDE, some
probability mass is assigned to empty portions of the support
(relative to $N$), thus leading to an overestimate of $p$ for some
portions of the support when $N$ is small. In contrast the histogram
estimator assigns zero probability mass for all bins with no data
points, thus leading to an underestimate of the probability mass of a
given bin. Because of these opposing properties, KDE and histogram
estimates will, for small $N$, differ in the presence of strong sample
size effects; in particular, the KDE will often yield a lower TDMI
fixed point bias estimate than that of the histogram estimator. Thus
in the circumstance where $N$ is small, differences in the KDE and
histogram estimates of the TDMI fixed point identify the existence,
and quantify the contribution, of small sample size effects on the
bias estimate. Finally, when applying this technique, it is important
to take care that the bin width and bandwidth are roughly equivalent.


\section{Implementation of the fixed-point method for estimating TDMI bias}
\label{sec:our_tdmi_estimation_scheme}

\subsection{Data sets}
\label{sec:data_sets}
 
The first data set we utilize is generated by the Lorenz equations
\cite{lorenz}; these ordinary differential equations were formulated
by Edward Lorenz via truncation of the Navier-Stokes equations for use
as a ``toy-model'' for studying atmospheric dynamics.  In particular,
the Lorenz equations are a simplified model of convection in a fluid
comprised, qualitatively, of three coupled ODEs, two of which detail
the time-evolution of amplitudes, and one which details the phase
relating those amplitudes. The Lorenz equations are given explicitly
by:
\begin{align}
\dot{x} =  \sigma (y-x) \\
\dot{y} =  x(\rho-z)-y \\
\dot{z} =  xy - \beta z
\end{align}
where $\sigma = 10$ (Prandtl number), $\rho=-28$ (Rayleigh number),
and $\beta=\frac{8}{3}$.  The reasons we chose to use data generated
by the Lorenz equations include: (i) the dynamics have been
extensively studied and are well understood computationally,
geometrically, and statistically; (ii) the Lorenz equations have
``weakly periodic,'' yet chaotic dynamics similar to diurnal variation
in human beings and thus useful for comparison with data set two; and
(iii) Lorenz equation dynamics occur on multiple time-scales. Note
that we integrate the Lorenz equations with a standard Runga-Kutta
fourth order integration scheme with a step-size of $10^{-3}$. The
Lorenz time-series is then sampled once per $100$ steps.

\begin{figure}[tbp]
  \epsfig{file=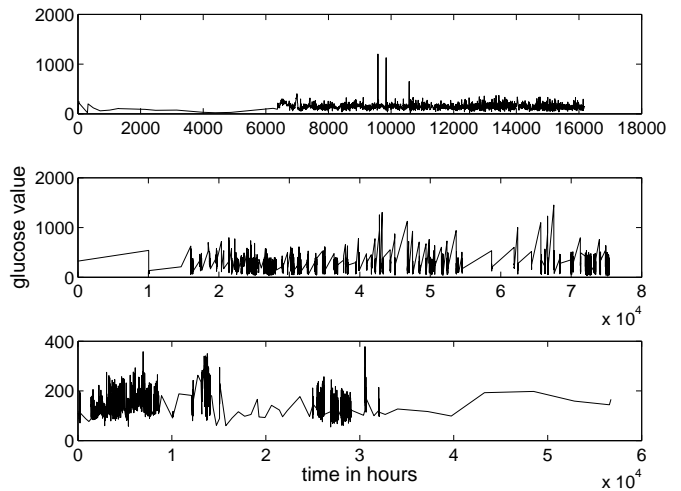, height=6.5cm}
  \caption{Time series for three well sampled patients (from top to
    bottom) with $\sim 2$, $\sim 8.5$ and $\sim 6.5$ year records
    respectively.  Note that the missing measurements have been
    interpolated by straight lines. The x-axis units are in hours; the
  y-axis units are in milligrams per decilitre (mg/dL).}
\label{fig:patients_glucose}
\end{figure}

The second data set we utilize includes three patients from the
Columbia University Medical Center (CUMC) patient database, containing
the records of $2.5$ million individuals over the course of $20$
years. We selected these patients by first narrowing the set of
patients to those who had glucose time series with at least $1,000$
points so that we could test the bias estimate techniques on the same
patient over a wide range of time series lengths. This restricted our
population to a few hundred patients. We then calculated the TDMI on the
each patient's full time series as well the first $100$ points of the
respective patient's time series. After examining a few hundred of
such cases, we selected three patients who were particularly
representative of the population. The three patients' time series of
glucose values are shown in Fig. \ref{fig:patients_glucose}, thus
displaying the sampling rates, proportion of missing values
(which are interpolated as straight lines), and overall clustering and
irregularity of measurement times. In particular, notice that: patient
one has dense and somewhat uniform sampling over much of the $\sim 2$
year long record; patient two is more irregularly sampled than patient
one, but remains relatively uniformly sampled over the $8.5$ years of
the record; and patient three has bursts of measurements followed by
large gaps between measurements over the $\sim 6.5$ year long
record. From this it is clear that not all patients will be able to
resolve all time scales.  



\subsection{TDMI estimation results}
\label{sec:tdmi_estimtion_results}

First, it is important to understand how the TDMI behaves (e.g., what
it can resolved, and to what level of accuracy) as $\delta t$ is
varied for time series of differing lengths. But, because the accuracy
of the TDMI estimate depends fundamentally on the number of pairs of
points within a time series that are separated by a given $\delta t$,
and because the cardinality of this set depends on both the length of
the time series and the relative density of sampling, it can be
difficult to understand how to apply the TDMI to a time series. Thus,
first we will consider how the TDMI behaves in the context of various
time series before considering how the TDMI varies with sampling
frequencies and sampling irregularities. Similarly, before considering
the ensemble averages of TDMI bias estimates, it is important to
understand the variation and variance of these estimates. In
particular, it is important to visualize how the TDMI, as a measure of
correlation decay, decays to the infinite time asymptotic state that
is used in the random permutation bias estimate. Note that for this
subsection we reduce the size of the data set from $N$ to $M$ by
taking the first $M$ time points in the time series.

\subsubsection{Lorenz equations results}
\begin{figure}[tbp]
  \subfigure[Long $\tau$ TDMI for Lorenz equation data with $100,000$ points]{ 
    \epsfig{file=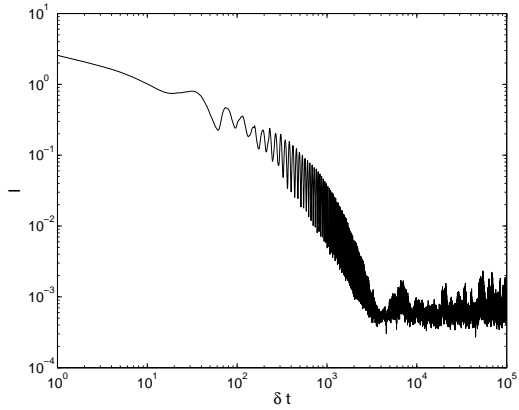, height=5.5cm}
    \label{fig:lor_tdmi_all_pts_a}
  }
  \subfigure[TDMI for Lorenz equation data with $10,000$ points]{ 
    \epsfig{file=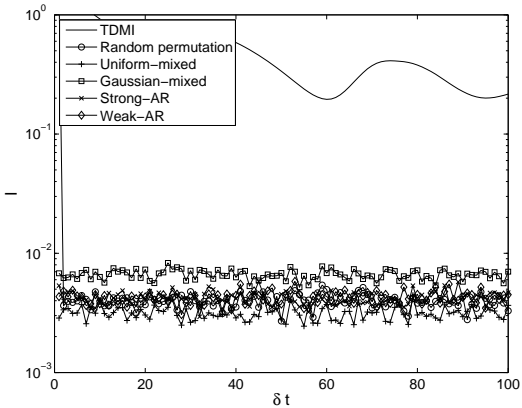, height=5.5cm}
    \label{fig:lor_tdmi_all_pts_b}
  }
  \subfigure[TDMI for Lorenz equation data with $1,000$ points]{ 
    \epsfig{file=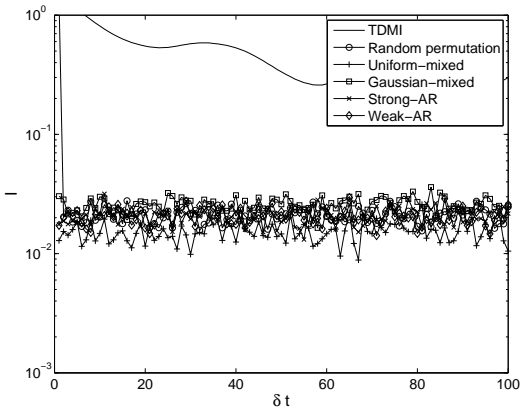, height=5.5cm}
    \label{fig:lor_tdmi_all_pts_c}
  }
  \subfigure[TDMI for Lorenz equation data with $100$ points]{ 
    \epsfig{file=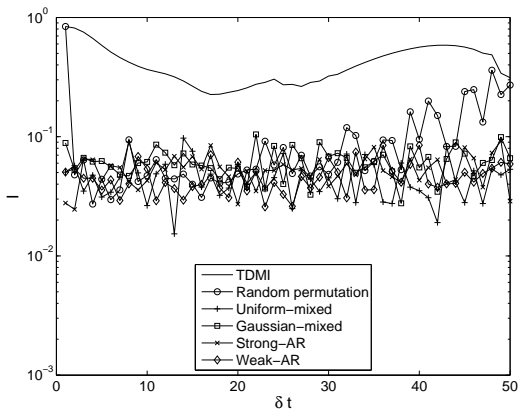, height=5.5cm}
    \label{fig:lor_tdmi_all_pts_d}
  }
\caption{Time-delay mutual information estimates and bias for the
  Lorenz equation data strings of
  various lengths.}
\label{fig:lor_tdmi_all_pts}
\end{figure}

Figure \ref{fig:lor_tdmi_all_pts} details the TDMI and bias estimates
for the Lorenz equations.  As can be seen in the long-time plot,
nonlinear correlations do not completely dissipate until measurements
are separated by at least $1,500$ data points at the given measurement
frequency.  The raw TDMI estimates for the both the $1,000$ and
$10,000$ point strings are identical, certainly up to the bias and
error.  The TDMI results for the $100$ point string, while a
qualitatively different raw TDMI graph from the $1,000$ and $10,000$
point estimates, are not wildly different; moreover, the raw TDMI
estimated on the $100$ point data set is not on the same order of
magnitude as any of the bias estimates.  Thus, while $100$ points may
not be enough for an extremely accurate TDMI estimate, it is enough to
establish a difference between persistent, existent correlation and
bias.  The variance of the various bias estimates again appears to
follow a power-law in the number of points in a data string ---
implying that small sample size effects do not dominate the estimate
of the estimator bias.  And finally, the different bias estimation
techniques do not differ enough to cause any difference in
interpretation of the TDMI signal; however, the Gaussian-mixed
estimate in the presence of long data-strings appears to be an upper
bound on other bias estimates.

\subsubsection{Glucose results}

\begin{figure}[tbp]
  \epsfig{file=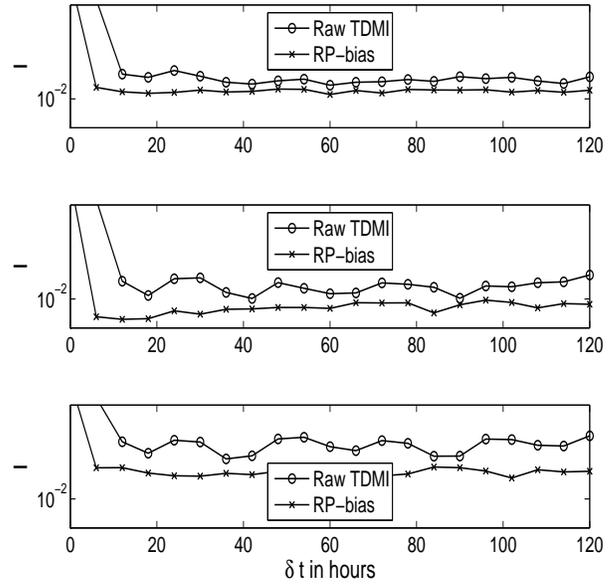, height=8cm, width=8cm}
  \caption{Time-delay mutual information and bias for delays of $0$ to
    $120$ hours ($5$ days) for three well sampled patients. Notices:
    the bias has low variance and is below the raw TDMI curve; the
    diurnal peaks in predictability; the variation between heights of
    the respective patients' TDMI peaks.}
\label{fig:glucose_all_patients_tdmi_bias}
\end{figure}

To show that the TDMI, and our bias estimates are not sensitive to
sampling frequency or specific source, consider
Fig. \ref{fig:glucose_all_patients_tdmi_bias}, where the TDMI and
random permutation bias is plotted for the three patients. In all
cases the bias is below the TDMI signal as expected.  Moreover, the
bias estimates show relatively little variance, implying that the
\emph{distribution} of bias estimates will be a rather peaked
distribution. Note that the each patient has a daily peak in
predictability, meaning that points separated by $24$ hours are the
most predictive of one another aside from points separated by less
than $6$ hours. However, not all patients have the same magnitude of
TDMI peak, meaning, the patients are not all equally
predictable. While we hypothesize that this variance in the TDMI daily
peaks is related to the relative health of the given patient's
endocrine system, a more detailed discussion is beyond
the scope of this paper. 

\begin{figure}[tbp]
  \subfigure[TDMI for glucose data with $\sim 4000$ points]{ 
    \epsfig{file=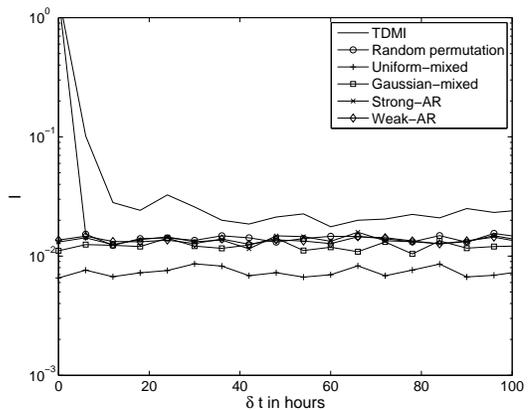, height=5.5cm}
    \label{fig:glu_tdmi_all_pts_a}
  }
  \subfigure[TDMI for glucose data with $1,000$ points]{ 
    \epsfig{file=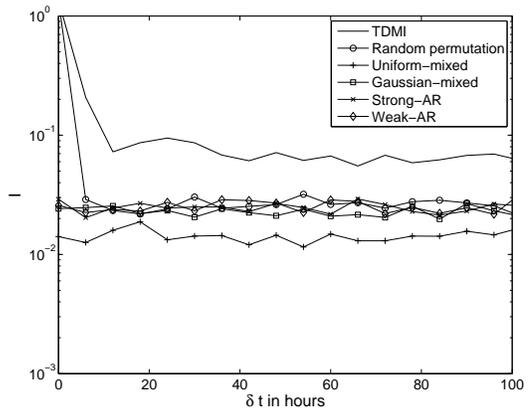, height=5.5cm}
    \label{fig:glu_tdmi_all_pts_b}
  }
  \subfigure[TDMI for glucose data with $100$ points]{ 
    \epsfig{file=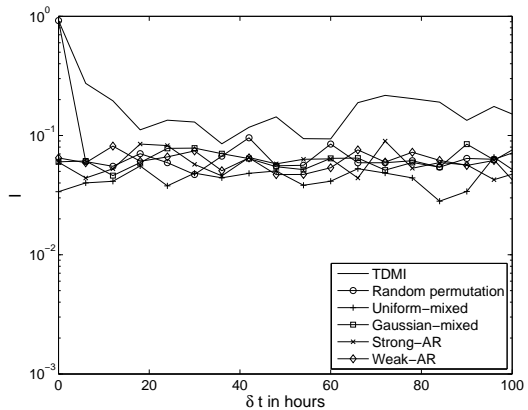, height=5.5cm}
    \label{fig:glu_tdmi_all_pts_c}
  }
  \subfigure[TDMI for glucose data with $20$ points]{ 
    \epsfig{file=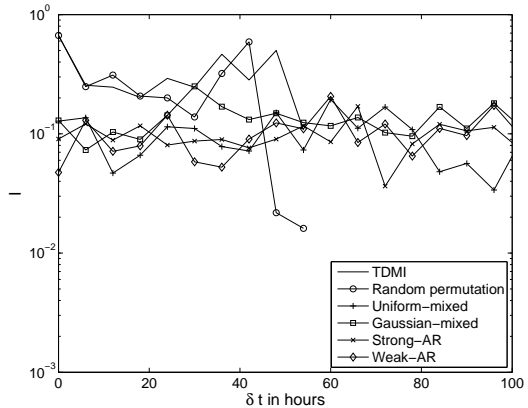, height=5.5cm}
    \label{fig:glu_tdmi_all_pts_d}
  }
  \caption{Time-delay mutual information estimates and bias for
    patient one's glucose time series of various lengths.}
\label{fig:glu_tdmi_all_pts}
\end{figure}

Recall that our goal is to understand how to estimate bias of the TDMI
algorithm in a fast, reliable way based on the data. Therefore, to
demonstrate and test the TDMI bias estimation algorithms, will focus on
patient one, the patient whose TDMI signal is the most difficult to
resolve. Figure \ref{fig:glu_tdmi_all_pts} details the TDMI of the
glucose values for patient one.  First notice that the $\sim 4,000$
and $1,000$ point TDMI results are very similar, including the
correlation peak at $24$ hours; however, the $\sim 4,000$ point TDMI
appears to resolve a $48$ hour peak as well, implying that the number
of points does effect the detail of resolution of a given signal.
Moreover, in both the $\sim 4,000$ point and the $1,000$ point TDMI
calculations, the TDMI signal is never buried in the
bias. Nevertheless, there is an important difference between the
glucose results and the Lorenz-based results: for the glucose-based
results, in general the difference between the TDMI and the bias is
smaller when $N$ is larger ($I(4000)-B(4000) < I(1000)-B(1000)$)
whereas for the Lorenz-based results the opposite is true.  Said
simply, the Lorenz-based result produces the expected outcome, whereas
the glucose-based result produces a surprise outcome. There are two
points to make about the unexpected nature of the glucose-TDMI results
for the $4,000$ and $1,000$ data point data sets. First, while the
difference between the TDMI and the bias is smaller when $N$ is
larger, this effect is relatively small and does not change any of the
conclusions that can be drawn from either TDMI signal.  Second, it is
unlikely that this data source is stationary; in particular, this
patient had a terminal disease and was slowly failing, and thus we
hypothesize that the amount of predictive information in this
patient's signal was likely decreasing (we will not attempt to further
justify this hypothesis here). Refocusing on the bias estimates for
the $1,000$ and $4,000$ point cases, overall, the various bias
estimates yield very similar results (all bias estimates are on the
same order of magnitude).  Nevertheless, note that here the
uniform-mixed is slightly separated from the rest of the bias
estimates, whereas for the other examples the Gaussian-mixed was
separated from the other estimates.

In contrast to the higher-$N$ cases, the TDMI estimate for the $100$
point data string rises after $\delta t \approx 48$ or equivalently
after two days, and moreover differs qualitatively from the $1,000$
and $\sim 4,000$ point TDMI estimates.  Thus precise TDMI estimates
beyond two days with a resolution of quarter days are unlikely with
only $100$ points even if the points are sampled relatively evenly in
time.  Nevertheless, the TDMI estimated on the $100$ point data set
shows a sharp decay in correlation over the first $48$ hours.  So,
while the TDMI estimated with only $100$ points will not resolve
perfectly, it will nevertheless retain some gross qualitative features
of the higher point analogs.  Finally, TDMI estimated using a $20$
point data set can resolve nothing; even the raw TDMI is buried in the
bias estimates and correlation decay over even the first $12$ hours is
not present.

\subsection{Comparing the fixed point bias estimation methods}
\label{sec:comparing_TDMI_methods}

Bias estimates are represented by distributions that canonically have
at least a mean and variance. Moreover, recall that the bias is a function of
the number of points in the sample set and the distribution of those
points (this controls the number of empty bins which influences
bias). The features of a time series that control the number of points
in a given $\delta t$ bin are the length of the time series and the
sampling frequency (i.e., it is impossible to resolve a $\delta t =
10^5$ TDMI with a time series whose maximum time separation is
$10^4$). Therefore, in an effort to minimize parameter variation, we
will not change the sampling rate and merely change the length of the
time series (they have the identical effect, reducing the number of
points in given $\delta t$ bins).

\begin{figure}[tbp]
  \subfigure[Mean and standard deviation for each fixed point bias
    estimate method for Lorenz equation data.]{ 
    \epsfig{file=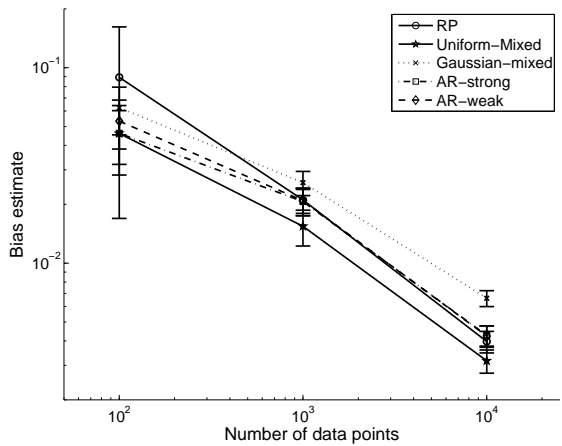, height=6cm}
    \label{fig:tdmi_compare_b}
  }
  \subfigure[Mean and standard deviation for each fixed point bias
    estimate method for glucose data.]{ 
    \epsfig{file=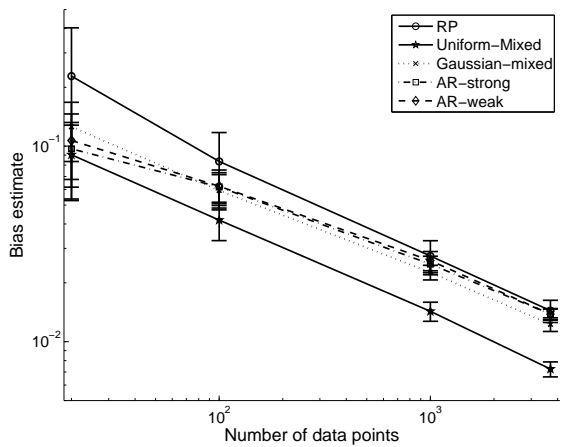, height=6cm}
    \label{fig:tdmi_compare_c}
  }
  \subfigure[Mean and standard deviation for each fixed point bias
    estimate method for glucose data for \emph{histogram and KDE
      estimators}; note the convergence of the methods as the number
    of points increases.]{ 
    \epsfig{file=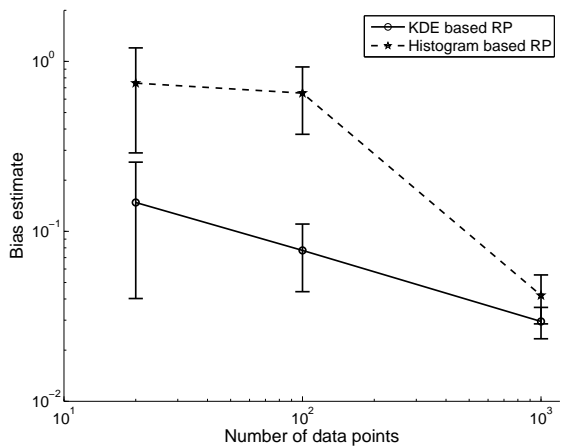, height=6cm}
    \label{fig:tdmi_compare_d}
  }
\caption{Comparing the bias estimation methods via both the mean and
  standard deviation of each bias estimate method
  (Figs. \ref{fig:glu_tdmi_all_pts_b} and \ref{fig:glu_tdmi_all_pts_c})
  and the estimator type (Fig. \ref{fig:glu_tdmi_all_pts_d}) as a
  function of the number of points in the glucose data set.}
\label{fig:tdmi_compare}
\end{figure}

With this in mind, we demonstrate how the permutation-method performs
by comparing the mean and variance of the bias estimation methods to
one another for different sample sizes, and for different estimators
in three steps. First, Fig. \ref{fig:tdmi_compare_b} represents the
variation of the bias estimates on a uniformly sampled time series
(Lorenz data) of length $N$; note that such a time-series will have
$N-\delta t$ pairs of points in each $\delta t$ bin. This figure thus
demonstrates how the different bias estimates behave as the absolute
number of points use for the PDF estimate varies. In practice, no one
characterizes a time series by the number of points separated by a
fixed time. Therefore, second, Fig. \ref{fig:tdmi_compare_c}
demonstrates the same calculation for the non-uniformly sampled
patient one (top plot, Fig. \ref{fig:patients_glucose}). Here again,
we do not alter the sampling rate which is already irregular, but
merely take the first $20$, $100$, $1,000$, and $4,000$ points in the
time series. This figure demonstrates that the bias estimate methods
apply well on real, irregularly time series. Finally, because
statistical estimates are made with concrete statistical estimators,
Fig. \ref{fig:tdmi_compare_d} demonstrates how the bias estimation
methods compare for histogram and KDE PDF estimators.

\subsubsection{Summary of conclusions}

\textbf{The broad observations relevant to the KDE-based TDMI bias estimate
comparisons include:} \textbf{(a)} uniform-mixed bias estimate is usually the
lower bound or nearly the lower bound on bias estimates; \textbf{(b)} the
Gaussian-mixed bias estimate converges to the uniform-mixed bias
estimate in the limit of short data strings and is often an upper
bound on all bias estimates for long data strings; \textbf{(c)} the
Gaussian-mixed and uniform-mixed bias estimates have the least
variance over data string lengths with Gaussian-mixed bias estimates
being an upper bound on the uniform-mixed bias estimates; \textbf{(d)} the
AR-mixed bias estimates converge to the uniform-mixed bias estimates
in the limit of short data strings and often in the limit of a long
data strings where the AR-mixed weak bias estimate is always an upper
bound on the AR-mixed strong bias estimate; \textbf{(e)} the random-permutation
bias estimate progresses from a lower bound bias estimate for long
data strings, usually coinciding with the uniform-mixed bias estimate,
to an upper bound bias estimate on short data strings; \textbf{(f)} the fixed
point bias estimates in all cases roughly obey the $\frac{1}{N}$
dependence on the number of points shown analytically for the
histogram estimator; and \textbf{(g)} all of the fixed point bias estimation
techniques yield estimates on the same order of magnitude.

\subsubsection{Detailed support of conclusions}

Observation (a), that the uniform-mixed is always a lower bound on the
bias estimates, is expected in light of how information entropy is, in
general, maximized \cite{jaynes_stat_mech_I}.  The uniform-mixed bias
estimate fits only a single parameter, the mean, to estimate the
distribution of the data.  Thus the uniform-mixed bias estimate
assumes the least possible information regarding the
intra-distribution correlations.  Said differently, all correlations
between subsets of the domain of the distribution are weighted
uniformly, thus minimizing the bias estimates in most circumstances.
Similarly, observation (b), that the Gaussian-mixed bias estimate
converges to the uniform-mixed bias estimate in the limit of few
points is largely due to the fact that the Gaussian-mixed bias
estimate is dependent on both the mean and variance of the target
distribution.  Because the variance for distribution with very few
points is very similar to the width of the domain of the target
distribution, the Gaussian-mixed and uniform-mixed bias estimates
nearly coincide.  That the Gaussian-mixed bias estimate is an upper
bound on other bias estimates for long data strings is related to the
fact that the variance of the Gaussian was fixed at one and the
entropy of a Gaussian is proportional to the variance.  Finally, that
the uniform-mixed and Gaussian-mixed bias estimates have the least
variance over the length of the data strings is largely due to the
fact that their respective marginals are completely independent of the
distribution of data.  That the AR-mixed bias estimates differ in this
regard is due to the fact that the estimated distribution that the
random numbers are generated from depends on each data set.

Observation (d), that the AR-mixed bias estimates converge to the
uniform-mixed bias estimates in the limits of long and short data
strings while remaining below the Gaussian-mixed bias estimates for
data strings of intermediate lengths has a simple explanation.
In particular, for long data strings, the AR-mixed bias estimate
converges to the random-permutation bias estimate because there is
enough data for the AR-mixed distribution to reasonably approximate
the real data distribution.  This essentially verifies that the fixed
point method of bias estimation we propose in this work is reasonable
because the TDMI estimates from fabricated data from a distribution
that resembles the distribution of the original data set coincides
with the TDMI results from the randomly permuted data.  For short data
strings, the AR-mixed bias estimate converges to the uniform bias
estimate because with so few points, the AR-mixed distribution that
fits the data best is often a uniform distribution as per the
principle of maximum entropy.

Observation (e), that the random-permutation bias estimate works well
for long data strings and poorly for short data strings, is the
result of most practical importance because it implies that using a
randomly permuted bias estimate for short data strings may yield an
overestimate of the bias. In particular, here when we randomly permute
the data string, the permutation is executed \emph{without
  replacement}, meaning that the data values are randomly permuted
only, never replaced.  This means that for a randomized data string of
length $n$ (or $\delta t_{max}$), the average time separation between
permuted values will be $\frac{n}{3}$ (or $\frac{\delta t_{max}}{3}$).
Thus, when $n$ is small (say, $n=10$), the randomly permuted data set
can have a TDMI bias estimate that is actually larger than the raw
TDMI estimate because on average, the data pairs are closer in time on
average, than for the raw data string.  This implies that using a
random permutation bias estimate for short data strings will at best
grossly overestimate the bias, and at worst will give a bias estimate
that will be unusable (e.g., greater than the raw TDMI value).  This
is the primary reason why we have proposed using the data fabrication
techniques for estimating TDMI bias for sparse time series.

Observations (f) and (g) imply that the fixed point bias estimation
techniques are both robust in that there is not significant variation
across the four techniques and consistent with other analytical
estimates of the estimator bias despite the fact that the analytical
estimates of bias were calculated for different estimator. In
particular, that the bias estimates for KDE and histogram estimators
both have the $1/N$ bias dependency is expected given that in the bias
estimates between two independent random variables have the $1/N$ bias
dependency for both KDE and histogram estimators. Moreover, that all
the estimation techniques produce roughly the same results is more
surprising and implies that the number of points used for the estimate
is the most important variable for the fixed point bias estimation
technique. This means that the technique is insensitive to how the
time-dependency is removed.  Finally, that the fixed point bias
estimation techniques are similar to the analytical bias estimates
implies what is likely the most important feature: the fixed point
bias estimates do approximate the estimator bias accurately.

The estimator-dependent effects on the bias estimates that can be read
off Fig. \ref{fig:tdmi_compare_d} include: (i) for small $N$ there is
a substantial difference between the KDE and histogram TDMI estimates
whereas for large $N$ they agree; (ii) the histogram estimate is
greater than the KDE estimate and has a greater variance (i.e., the
histogram TDMI estimate approaches from above while the KDE TDMI
estimate approaches from below); (iii) the TDMI fixed point bias
estimate of both estimators decreases roughly proportional to $1/N$;
and (iv), the difference between the small and large $N$ TDMI fixed
point bias estimate is much smaller for the KDE-based TDMI estimate
than for the histogram-based TDMI estimate. This all implies that both
the KDE and histogram estimators can be used to estimate the estimator
bias in a TDMI calculation; for small sample sizes, the KDE-based
calculation appears to work better. Based on how KDE versus histogram
estimation schemes work, these results are not
surprising. Nevertheless, \emph{because there is an estimator
  dependence on the bias estimate that depends on sample size},
comparing results from two estimators can reveal small sample size
effects.

The reason why the difference between the KDE and histogram bias
estimates can be used to reveal small sample size effects lies in the
difference in how they assign probabilities to bins without points in the
situation where few bins have any points. The histogram estimator
underestimates the probability of bins with few points as it
assigns a strict zero if the bin contains no pairs. Thus, histogram
estimators, in the presence of small sample sizes, tend to represent
sharply peaked distributions.  In contrast, the KDE estimator weights
each bin (or point relative to its bandwidth) more like a uniform
distribution in the presence of few samples. Thus, the KDE estimator
tends to represent too smooth a distribution, or a uniform
distribution. Therefore, KDE and histogram estimators approach the
infinite $N$ bias limit from below and above respectively. 

\section{Summary and discussion}

\textbf{Robust and consistent with previous bias estimate
  techniques}---Based on the analysis in the previous section, we
claim that the \emph{fixed-point bias estimation method} for
estimating the TDMI bias is accurate, easy to use, robust with respect
to differing methodologies, and works for a variety of estimators.  In
particular, all the time-dependency removal schemes we employed
qualitatively reproduced the $1/N$ bias dependency previously know for
histogram estimators and observed in the KDE estimators. Moreover, all
the bias estimate schemes we employed are easy to use and compute as
fast as the standard TDMI computation. Finally, because the bias
estimation technique relies on re-ordering or replacing the data
\emph{before the PDF estimates are carried out}, this estimation
scheme will work for all PDF estimators.

\textbf{Contrasting the bias estimate methods}---Given data strings
that have a reasonable density and length, the random permutation bias
estimate will likely yield both the best estimate of the bias and the
fastest implementation of a bias estimate.  Nevertheless, for short
data strings, the random-permutation bias estimate is likely to
overestimate the bias. This overestimate occurs primarily because
randomly permuting a few time-correlated points will, on average,
force an average maximum time separation of one-third the length of
time represented in the data set. For short data-sets, this can imply
a significant amount of time-based correlation. Luckily, in the
context of short data strings, the uniform-mixed bias estimates do
work reasonably well.  Moreover, in all data-string length contexts,
using more sophisticated means of fabricating data such as
accept-reject methods for generating distributions do not, in general,
outperform either the uniform-mixed or Gaussian-mixed, or random
permutation bias estimates.  Because these more sophisticated bias
estimation techniques can be significantly more computationally
intensive, we do not recommend their implementation.

\textbf{Data-based constraints on the TDMI calculation}---For data sets
with fewer than $50$ points, very little TDMI related information can
be gained.  However, for data sets with as few as $100$ points, very
often qualitative time-based correlation information such as a simple
decay in correlation can be determined.  As the number of points
increase from $100$ up to $1,000$ and beyond, the TDMI and TDMI bias
calculations continue to improve.  For the data sets we considered in
this paper, beyond $1,000$ points all of the qualitative and some
quantitative conclusions drawn from the TDMI signal did not change
when more data points were added.  Nevertheless, these statements
depend on the density of the data in time, the time-resolution
desired, and the uniformity of the measurements in time. Finally, when
there is concern about the presence of small sample size effects, a
comparison (via the difference) between histogram and KDE-based
estimates of the TDMI fixed point bias can be used to detect and
quantify the existence of the small sample size effects.

\textbf{Consequences for data-set aggregation techniques}---Finally,
because little TDMI information can be gleaned from data strings of
lengths shorter than $100$ points, and because sparse time series may
have fewer than $100$ points, the usefulness of utilizing such sparse
data for statistical analysis will often hinge on the ability to
aggregate like time series into a single long time series. With
respect to the TDMI and its bias calculation, for medical data where
most patients are similar statistically and thus may be allow for
aggregation, but for which the patients all have few points and their
individual bias contributions must be handled within patient, it
may be best to use fabricated data to estimate the TDMI bias.  Or,
said differently, for data sets that allow for aggregation, if the
bias must be estimated on an intra-string basis, random-permutation
bias estimates will likely overestimate the bias.  In such cases, we
recommend calculating \emph{both} the random-permutation and the
uniform-mixed bias estimates.  Contrasting these two bias estimates
will likely yield a fruitful interpretation and estimate of the bias.


DJA and GH would like to acknowledge the Columbia University
Department of Biomedical Informatics data-mining group for helpful
discussions; D. Varn and J. F. S. Dias for a careful reading of the
manuscript; and financial support provided by NLM grant RO1 LM06910,
an award from Microsoft Research for the Phenotypic Pipeline for
Genome-wide Association Studies, and a grant from The Smart Family
Foundation.


\end{document}